\documentclass[prb,reprint,showpacs,amsmath,amssymb,superscriptaddress]{revtex4-1}
\usepackage{graphicx}
\usepackage{color}
\usepackage[colorlinks,bookmarks=false,citecolor=blue,linkcolor=red,urlcolor=blue]{hyperref}
\usepackage{multirow}

\begin{document}
\title{Phase Diagram of Bosons in a 2D Optical Lattice with
  infinite-range Cavity-mediated Interactions }

\author{T. Flottat}
\affiliation{Universit\'e C\^ote d'Azur, CNRS, INLN, Valbonne, 06560, France}
\author{L. de Forges de Parny}
\affiliation{Laboratoire de Physique, CNRS UMR 5672,  \'Ecole Normale
  Sup\'erieure de Lyon, Universit\'e de Lyon, 46 All\'ee d'Italie,
  Lyon, F-69364, France} 
\affiliation{Physikalisches Institut, Albert-Ludwigs Universit\"{a}t
  Freiburg, Hermann-Herder Stra{\ss}e 3, D-79104, Freiburg, Germany}
\author{F. H\'ebert}
\email[Corresponding author: ]{frederic.hebert@unice.fr}
\affiliation{Universit\'e C\^ote d'Azur, CNRS, INLN, France}
\author{V.$\,$G.  Rousseau}
\affiliation{Physics Department, Loyola University New Orleans, 6363
  Saint Charles Ave., LA 70118, USA} 
\author{G.$\,$G. Batrouni}
\affiliation{Universit\'e C\^ote d'Azur, CNRS, INLN, France}
\affiliation{MajuLab, CNRS-UNS-NUS-NTU International Joint Research
  Unit UMI 3654, Singapore} 
\affiliation{Centre for Quantum Technologies, National University of
  Singapore; 2 Science Drive 3 Singapore 117542} 

\begin{abstract}
  High-finesse optical cavity allows the establishment of long-range
  interactions between bosons in an optical lattice when most cold
  atoms experiments are restricted to short-range interactions.
  Supersolid phases have recently been experimentally observed in such
  systems.  Using both exact quantum Monte Carlo simulations and
  Gutzwiller approximation, we study the ground state phase diagrams
  of a two-dimensional Bose-Hubbard model with infinite-range
  interactions which describes such experiments.  In addition to
  superfluid and insulating Mott phases, the infinite-range
  checkerboard interactions introduce charge density waves and
  supersolid phases. We study here the system at various particle
  densities, elucidate the nature of the phases and quantum phase
  transitions, and discuss the stability of the phases with respect to
  phase separation. In particular we confirm the existence and stability of a
  supersolid phase detected experimentally.
\end{abstract}


\maketitle

\section{Introduction}

In the past fifteen years, cold atoms and Bose Einstein condensates
have proven invaluable tools to study the physics of interacting
quantum systems \cite{Bloch_2008}.  Optical lattices
\cite{Greiner_2002, Bloch_2008} and Feshbach resonances
\cite{Chin_2010} have been used to drive these systems into strongly
interacting regimes. Recently, it has been shown that coupling between
cold atoms and an electromagnetic field, for example by putting a
condensate in a cavity, leads to interesting collective phases of
light and matter \cite{Ritsch_2013}. For example, the Dicke transition
\cite{Baumann_2010}, superradiant Mott phases \cite{Klinder_2015},
self structuration of atoms and light \cite{Labeyrie_2014}, and
crystallization \cite{Ostermann_2016} have been observed in such
systems. Many theoretical predictions have also been made such as the
possibility to observe Bose glasses \cite{Rieger_2013}, localization
\cite{Rojan_2016}, or synchronization of quantum dipoles
\cite{Zhu_2015}.

Here we are especially interested in a recent experiment by the
ETH-Zurich group \cite{Esslinger_2016} where a cloud of cold atoms is
placed in an optical lattice and inside an optical cavity. The field
of the cavity mediates an effective infinite range interaction between
the atoms which favours a density difference between neighbouring sites
of the optical lattice.  This experiment attracted a strong interest
as it provides one of the first observations of the elusive supersolid
phase \cite{Penrose_1956, Gross_1957}.  This exotic phase is
characterized by both long range phase coherence and spatial ordering,
i.e. simultaneous diagonal and off-diagonal long range orders.
In a recent experiment, the same group observed a supersolid phase
with a symmetry breaking of a continuous space invariance
\cite{Leonard_2016}.

The experimental system is well described by a conventional
Bose-Hubbard model with an additional infinite range interaction
\cite{Esslinger_2016}, which takes into account the effect of the
cavity field.  This model and similar ones have been mostly studied within the (static
and dynamic) mean-field theory\cite{Bakhtiari_2015, Donner_2016, Mueller_2016, Zhai_2016, Byczuk_2016};
only one study has employed an exact
quantum Monte Carlo method in the hard-core limit \cite{Rieger_2016}.
In the current state of the literature, exact studies in the strong
(but finite) interacting regime are still missing. 

In this work, we use exact quantum Monte Carlo simulations to
determine the phase diagram of this model for several particle
fillings, and elucidate the nature of the observed quantum phase
transitions.  We also compare our results with mean-field results,
obtained with an approach based on the Gutzwiller ansatz and classical
Monte Carlo simulations \cite{Paramekanti_2014, Roscilde_2017}.

The paper is organized as follows: The Hamiltonian and the methods
used are presented in Sec.~\ref{sec_2}. Section~\ref{sec_3} is devoted
to the discussion of the mean field phase diagrams whereas the exact
ground-state phase diagrams, obtained by using quantum Monte Carlo
simulations, are discussed in Sec.~\ref{sec_4}.  Finally, conclusions
and outlook are provided in Sec.~\ref{sec_5}.

\section{Hamiltonian and Methods}
\label{sec_2}

\subsection{Bose-Hubbard model with infinite-range checkerboard
  interactions} 
\label{sec2_subA}

We consider spinless bosons in a two-dimensional square optical
lattice inside a high-finesse optical cavity.  The particles can hop
between nearest neighbouring sites of the lattice and interact
repulsively on site.  An additional effect due to the external cavity
field generates an effective infinite-range interaction between
particles\cite{Esslinger_2016}.  Integrating out the effect of the
field, the system is then shown to be governed by a Bose-Hubbard model
with Hamiltonian \cite{Esslinger_2016}:
\begin{eqnarray}
\mathcal {  \hat H}   =&&   - t \sum_{\langle {\bf r,s} \rangle}
\left(  b^\dagger_{\bf r} b^{\phantom\dagger}_{\bf s} + {\rm H.c.}
\right ) + U_{s}  \sum_{{\bf r} \in e,o } \frac{n_{\bf r} \left (
    n_{\bf r}-1\right )}{2} \nonumber\\ 
 &&-   \frac{U_{l}}{L^2}   \left (    \sum_{{\bf r} \in e} n_{\bf r}
   - \sum_{{\bf r} \in o} n_{\bf r}     \right )^2.  
\label{BH_Hamiltonian}
\end{eqnarray}
The bosonic operator $b^ \dagger_{\bf r}$ ($b^{\phantom\dagger}_{\bf
  r}$) creates (annihilates) an atom at site $\bf r$ and $n_{\bf r}=
b^{\dagger}_{\bf r} b^{\phantom\dagger}_{\bf r}$ is the corresponding
number operator.  The indices $e$ and $o$ denote respectively even and
odd lattice sites.  The first term of the Hamiltonian is the kinetic
term describing tunnelling with amplitude $t$ between nearest neighbour
sites {\bf r} and {\bf s} defined on a square lattice of $L\times L$
sites with periodic boundary conditions.  The second term represents
the on-site repulsive interactions between the atoms with strength
$U_s>0$.  The third term describes the infinite-range interaction with
amplitude $U_l>0$ and favours imbalanced populations between even and
odd sites.  $\mu$ will denote the chemical potential for simulations
performed in the grand canonical ensemble (GCE).

The Hamiltonian has a $U(1)\times \mathbb{Z}_2$ symmetry, associated
with the mass conservation ($U(1)$ symmetry), times the Ising
$\mathbb{Z}_2$ symmetry between the even and odd checkerboard
sublattices.

\subsection{Methods}
\label{sec2_subB}

The approximate Gutzwiller Monte Carlo (GMC) \cite{Paramekanti_2014, Roscilde_2017}
approach is a numerical method built on the combination of both the
Gutzwiller ansatz and the classical Monte Carlo method with Metropolis
algorithm \cite{Metropolis_1953}. This results in a semi-classical
lattice field theory which preserves the $U(1)$ symmetry, which is an
advantage compared to some of the mean-field approaches conventionally
used.  This method also allows the reconstruction of correlation
functions on a finite lattice cluster. The Gutzwiller mean-field state
takes the form
\begin{equation}
  |\Psi({\bf f})\rangle = \bigotimes_{{\bf r}=1}^{L^2} |\psi_{\bf
    r}\rangle = \bigotimes_{{\bf r}=1}^{L^2} \left( \sum_{n_{\bf r}
      =0}^{n_{\rm max}} f^{(\bf r)}_{n_{\bf r}}|n_{\bf r}\rangle
  \right)~,  
\label{Gutzwiller_wave_function}
\end{equation}
where $|n_{\bf r}\rangle$ is the state with $n_{\bf r}$ particles on
site $\bf r$ and where we introduced a cut-off $n_{\rm max}$ on the
number of particles per site. The ensemble ${\bf f} = \{ f^{(\bf
  r)}_{n_{\bf r}} \}$ of the complex $f^{(\bf r)}_{n_{\bf r}}$
coefficients is then sampled with the Monte Carlo method  \cite{Paramekanti_2014, Roscilde_2017} which
is especially useful at finite temperature but can also be used
in the low temperature regime.

\label{sec2_subC}

The Hamiltonian is also simulated by using the stochastic Green
function algorithm (SGF) \cite{SGF, directedSGF}, an exact quantum
Monte Carlo (QMC) technique that allows simulations in the canonical (CE) or
grand canonical (GCE) ensembles of the system at finite temperatures,
as well as measurements of many-particle Green functions.

We treat $L\times L$ lattices with sizes up to $L=14$ and fix $t=1$ to
set the energy scale.  Large enough inverse temperatures allow to
eliminate thermal effects from the QMC and GMC results (we used
inverse temperatures $\beta t=2L$ for the QMC simulations and up to
$\beta t=10^4$ for the GMC simulations).
 
In particular we focus mainly on simulations at fixed density $ \rho =
\sum_{\bf r} \langle n_{\bf r} \rangle/L^2$.  $N= \rho L^2$ is the
total number of particles.  The phase coherence is captured by the one
body Green function,
\begin{eqnarray}
G({\bf R})  = \frac{1}{ 2L^2} \sum_{\bf r} \langle b_{\bf r}^{\dagger}
b_{\bf r+R}^{\phantom\dagger} + {\rm H.c.} \rangle 
\label{Gfunction}
\end{eqnarray}
and its Fourier transform $n({\bf k})$ is the density of particles
occupying the wave vector ${\bf k}$.  The condensate fraction, 
i.e. the fraction of particles occupying the ${\bf k}=0$ mode, is
given by $n({\bf k}=0) = \sum_{\bf R} G({\bf
  R})/N\label{Condensate_fraction}$.
 We also calculate
the superfluid density $\rho_s $ given, in the QMC algorithm, by
fluctuations of the winding number\cite{roy} $W$, $\rho_s = \langle
W^2\rangle/(4t\beta)$.  Finally, we also calculate the density-density
correlation
\begin{eqnarray}
D({\bf R}) = \frac{1}{L^2}\sum_{\bf r}\langle n_{\bf r} n_{\bf r+R}
\rangle 
\end{eqnarray}
and its Fourier transform, the structure factor $S({\bf k}) =
\sum_{\bf R} e^{i {\bf k.R}} D({\bf R})/L^2$.  We particularly focus
on $S(\pi, \pi)$ as we expect checkerboard phases to appear.

\section{Gutzwiller Phase Diagrams}
\label{sec_3}

\begin{figure}[!h]
\includegraphics[width=1 \columnwidth]{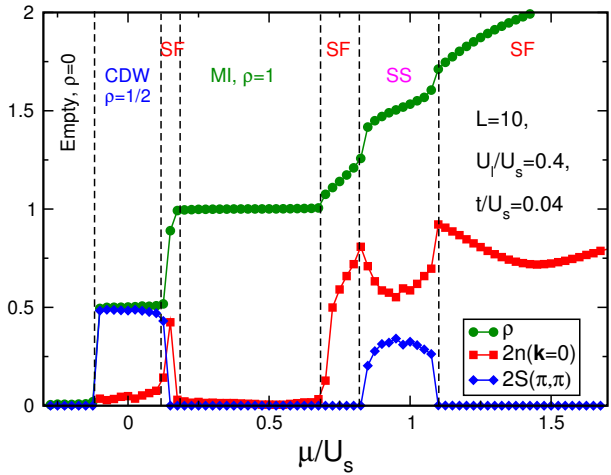}
\caption{(Color online) The density, $\rho$, condensate fraction, $n({\bf k}=0)$, 
and structure factor, $S(\pi,\pi)$, as functions of the chemical potential, $\mu$,
obtained with the GMC method at low temperature. We observe four phases:
solid with charge density wave (CDW), superfluid (SF), Mott insulator (MI), and
supersolid (SS).
\label{Figure_0}}
\end{figure}

\begin{figure}[h]
\includegraphics[width=1 \columnwidth]{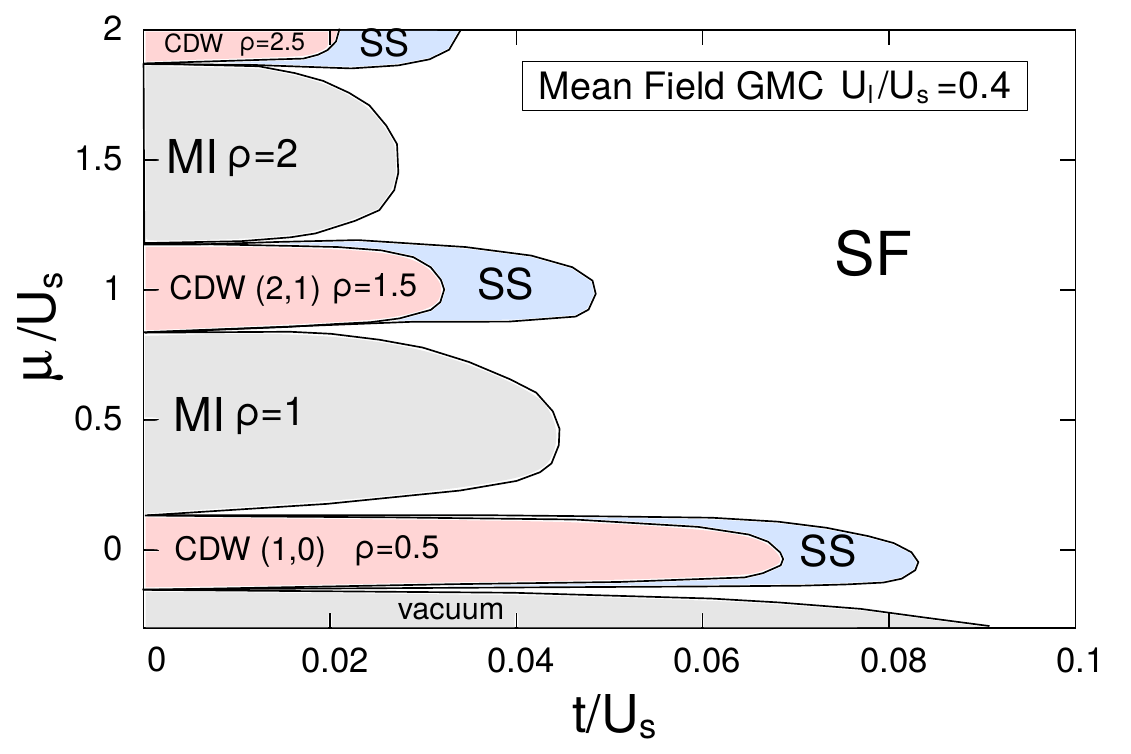}
\caption{(Color online) Ground-state mean field phase diagram in the
  plane $(t/U_s, \mu/U_s)$ obtained with the GMC method for
  $U_l/U_s=0.4$.  Four phases are observed: superfluid (SF), Mott
  insulator (MI), charge density wave (CDW) and supersolid (SS)
  phases.}
\label{Figure_2}
\end{figure}  
\begin{figure*}[t]
\hbox{\includegraphics[width=0.66 \columnwidth]{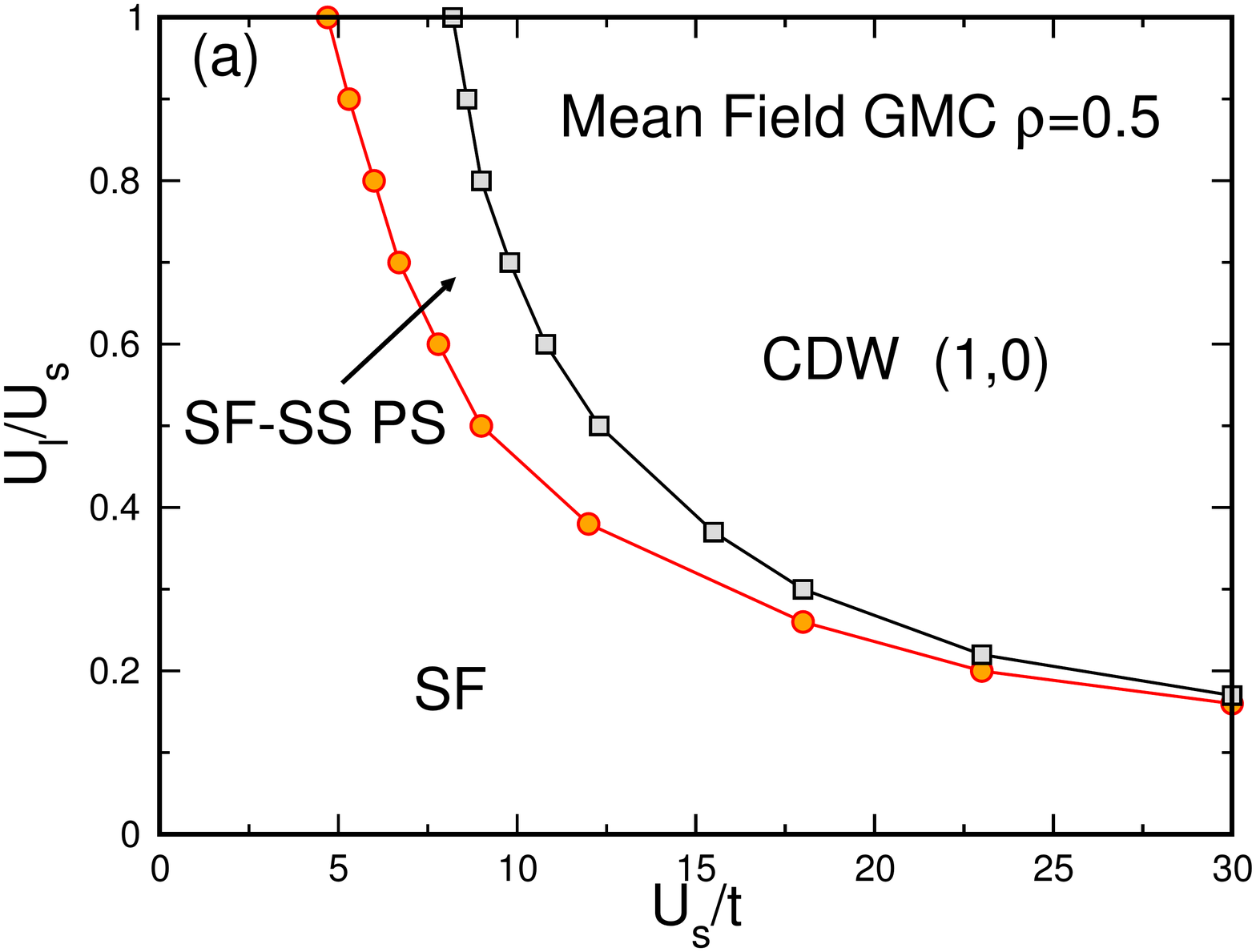}
  \includegraphics[width=0.66 \columnwidth]{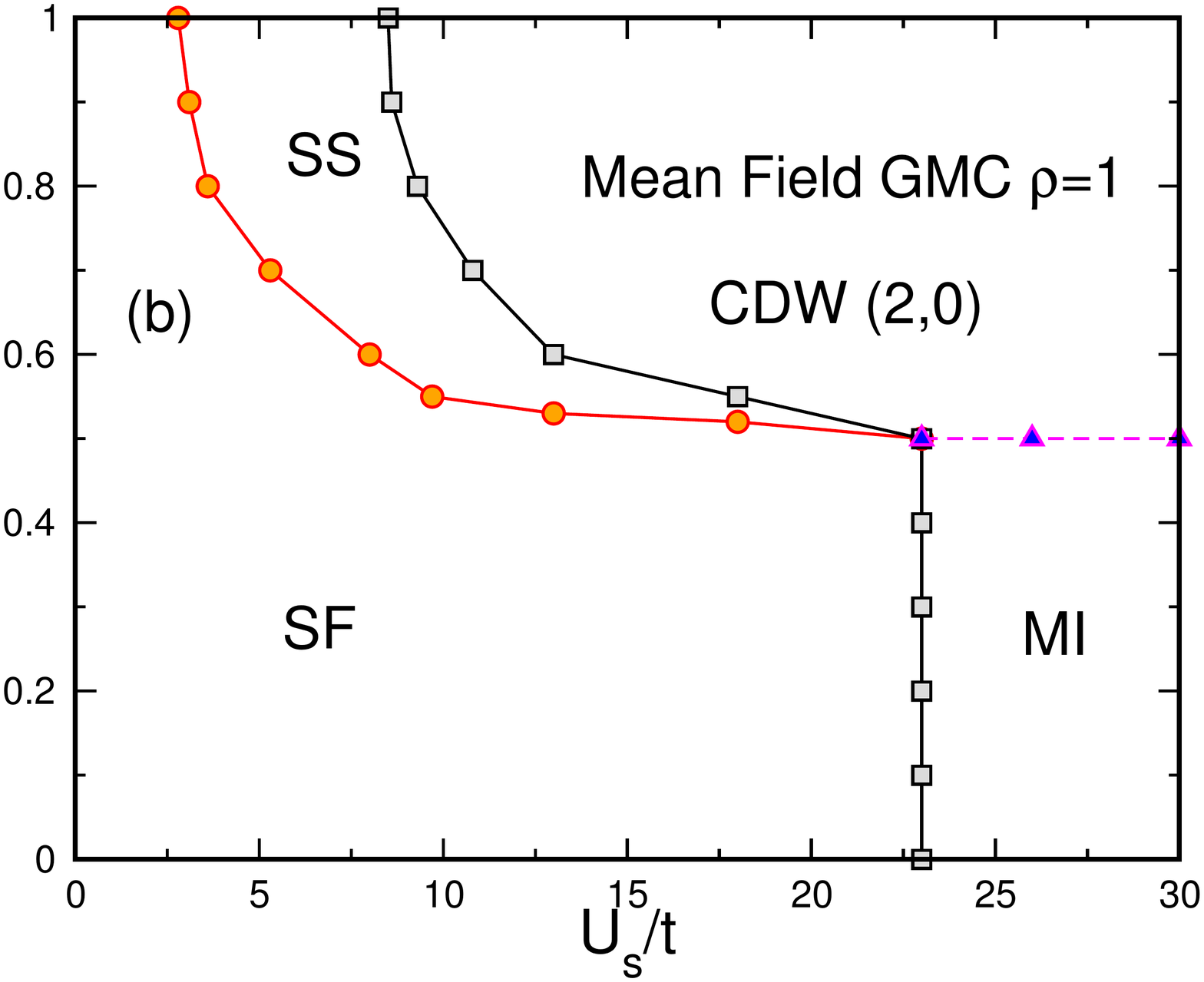}
\includegraphics[width=0.66 \columnwidth]{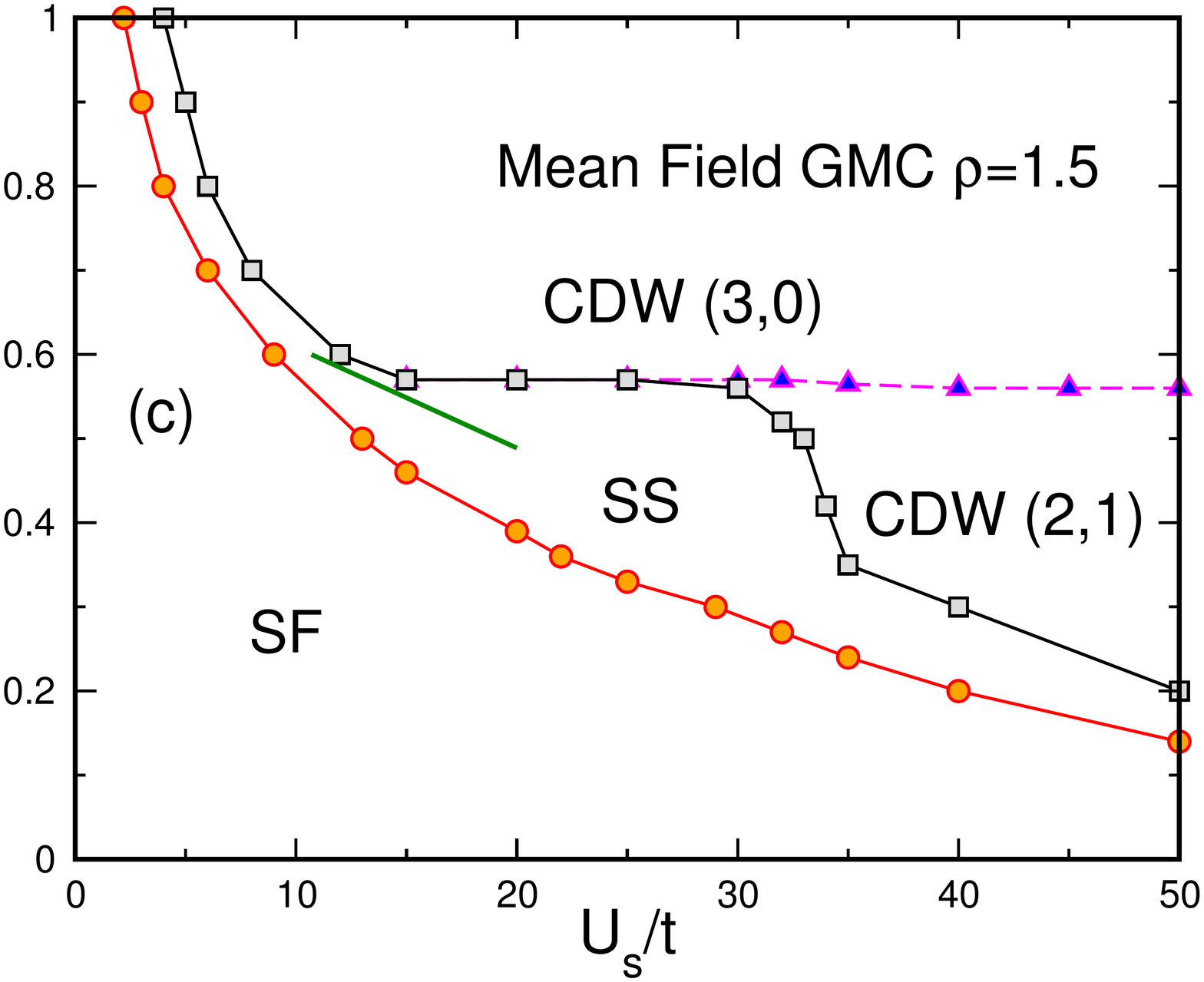}}
\caption{(Color online) Ground-state mean field phase diagrams
  obtained with the GMC method for three fillings (a) $\rho=1/2$, (b)
  $\rho=1$ and (c) $\rho=3/2$.  Four stable phases are observed: superfluid
  (SF), Mott insulator (MI), charge density wave (CDW) and supersolid
  (SS) phases.  The dashed lines indicate first-order transitions,
  otherwise the transitions are second order.  The green line in (c)
  corresponds to the cut through the SS phase used in
  Fig. \ref{figcrossover} (b).  
In Sec. \ref{substab}, we will discuss
  the stability of these phases and show that there
is a region of phase separation between a superfluid and a
supersolid (SF-SS PS) at $\rho=1/2$ instead of a SS phase.
\label{Figure_3}}
\end{figure*}

Since competing terms are involved in the Hamiltonian,
Eq.~(\ref{BH_Hamiltonian}), we expect four different phases at zero
temperature. For $U_l=0$, i.e. the standard Bose-Hubbard model,
it is well known that the competition between the kinetic and
interacting terms leads to two phases. Most of the phase diagram
consists of a Bose condensed (BEC) superfluid phase (SF) which
exhibits phase coherence indicated by $n({\bf k}=0) \neq 0$ and
$\rho_s \neq 0$. For integer particle densities, $\rho$, and strong
repulsion, $U_s$, there are also Mott insulating (MI) phases with
$n({\bf k}= 0)=0$ and $\rho_s = 0$.  Adding the checkerboard
interaction $U_l \neq 0$ offers the possibility to stabilize spatial
ordering, i.e. oscillations in the density signalled by $S(\pi,
\pi)\neq 0$.  In addition to the SF and MI phases -- for which $S(\pi,
\pi)= 0$ as the populations are balanced in these phases -- there is,
therefore, the possibility of two other phases: a charge density wave
(CDW) solid with vanishing coherence (for integer or half-integer
densities) and a supersolid (SS) phase exhibiting both spatial
ordering $S(\pi, \pi)\neq 0$ and phase coherence. In Fig. \ref{Figure_0},
we show $\rho$, $n({\bf k}=0)$, and $S(\pi,\pi)$ as functions of the chemical 
potential, $\mu$. We observe the four previously mentioned  phases.
The truncation $n_{\rm max}$ (up to $n_{\rm
  max}=8$) is chosen large enough for the results not to depend on it 
and the temperature is chosen low enough to be in the ground state
limit.

Using cuts such as Fig. \ref{Figure_0}, we determined
the phase diagrams in the $(t/U_s, \mu/U_s)$ plane for given values
of $U_l/U_s$ (see Fig.~\ref{Figure_2} for $U_l/U_s=0.4$).
The four phases, SF, MI, CDW, and SS, are clearly seen in this 
phase diagram.
In the zero hopping limit $t=0$, the CDW phases appear between the
Mott insulator phases, reducing the energy gap of the MI phases to
$\Delta = U_s- U_l$ for $0<U_l<U_s/2$.  At large hopping amplitude $t$
the system is in the SF phase.  In the intermediate regime, for
moderate hopping, we systematically observe SS phases at the tip of
the CDW lobes.  Note that this phase diagram is in a good agreement
with previous mean field studies \cite{Donner_2016, Mueller_2016}.
The SS-SF phase transition at the tip of the lobes is found to be
continuous, as observed in Ref. \cite{Donner_2016}, and not
first-order as observed in Ref.\cite{Mueller_2016}. 

As reported in Refs.~\cite{Donner_2016, Mueller_2016}, beyond
$U_l=U_s/2$, the nature of the phases change, as phases that show a
density modulation are favoured (Fig.~\ref{Figure_3}). The Mott phases are thus replaced with
CDW phases. In the small $t$ limit, the Mott phase at $\rho=1$ is
replaced with a phase where, depending on the way the symmetry breaks,
even (odd) sites are occupied by 2 bosons while odd (even) sites are
empty. This is what we call a CDW (2,0) phase.  The CDW phases at
half-integer fillings are also affected. For example, the CDW (2,1)
phase, i.e. having alternately doubly and singly occupied sites,
observed at $\rho=3/2$ below $U_l/U_s = 1/2$ is replaced with a (3,0)
phase. The supersolid region is also much larger as it surrounds
completely the CDW lobes for $\rho>1/2$.  The bosons are expected to
collapse for $U_l>U_s$ in $t=0$ limit
\cite{Mueller_2016}. 

We now focus on the phase diagrams in the $(U_s/t, U_l/U_s)$ plane
maintaining the density fixed by adjusting the chemical potential.
The GMC mean field phase diagrams for densities $\rho=0.5$, $\rho=1$,
and $\rho=1.5$ are plotted in Fig.~\ref{Figure_3}.  Our results are
similar to the phase diagrams of Ref.~\cite{Donner_2016}.  We observe
that the supersolid phase surrounds CDW phases, which means that it
only appears for $U_l/U_s>1/2$ for integer densities (as $\rho=1$)
while it is present for smaller $U_l$ at half-integer
densities. At $\rho=1/2$, what appears at first sight
to be a supersolid phase, where $n({\bf k}=0)$ and $S(\pi,\pi)$ are both
non zero, is not a stable phase but a region of the phase diagram
where we observe a phase separation between a superfluid with $\rho<1/2$
and a supersolid with $\rho>1/2$ (SF-SS PS region). We will discuss this point later
when we will study the stability of the different phases 
(Sec. \ref{substab}).

\section{Quantum Monte Carlo Phase Diagrams}
\label{sec_4}
\begin{figure*}[t]
\hbox{\includegraphics[width=0.66 \columnwidth]{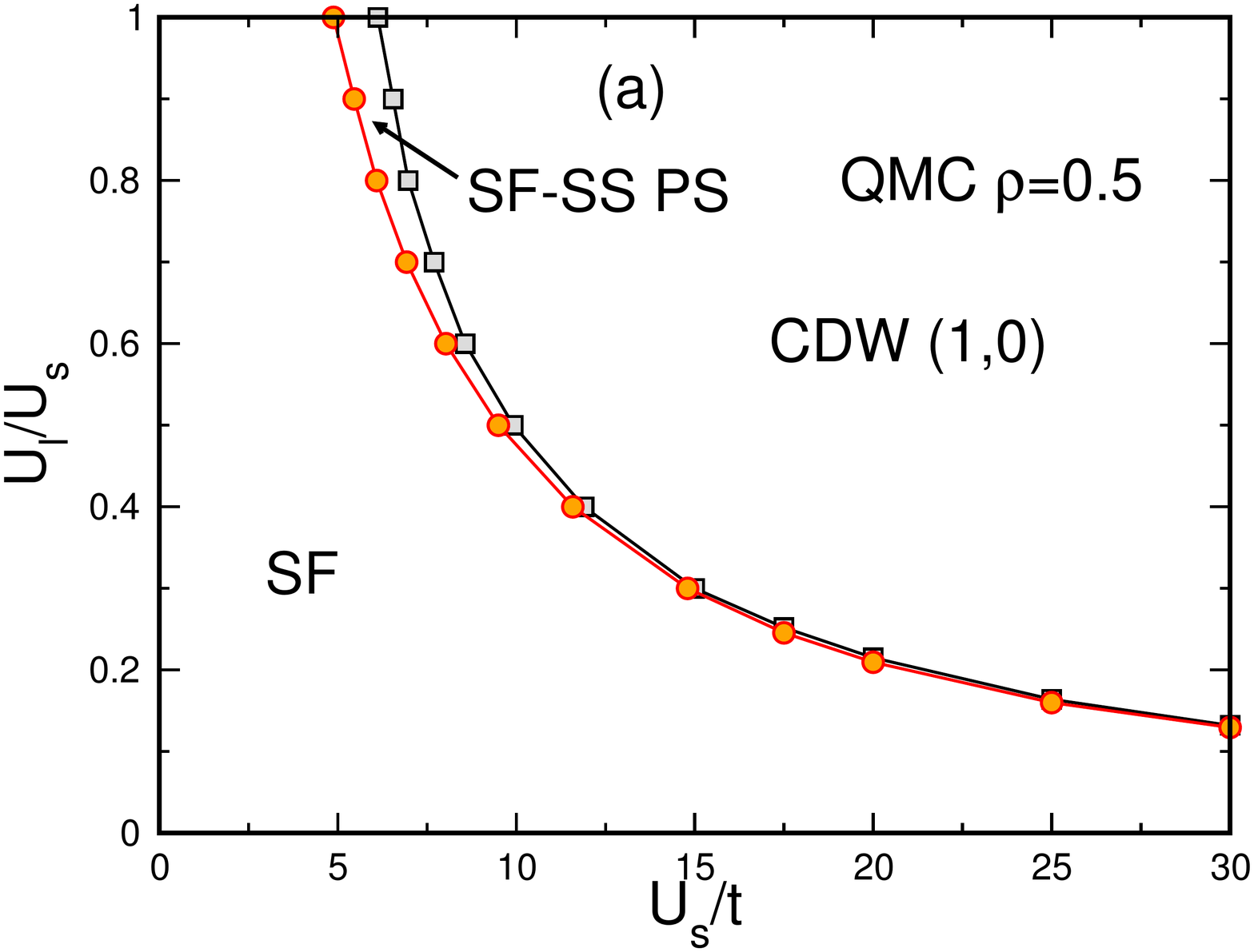}
\includegraphics[width=0.66 \columnwidth]{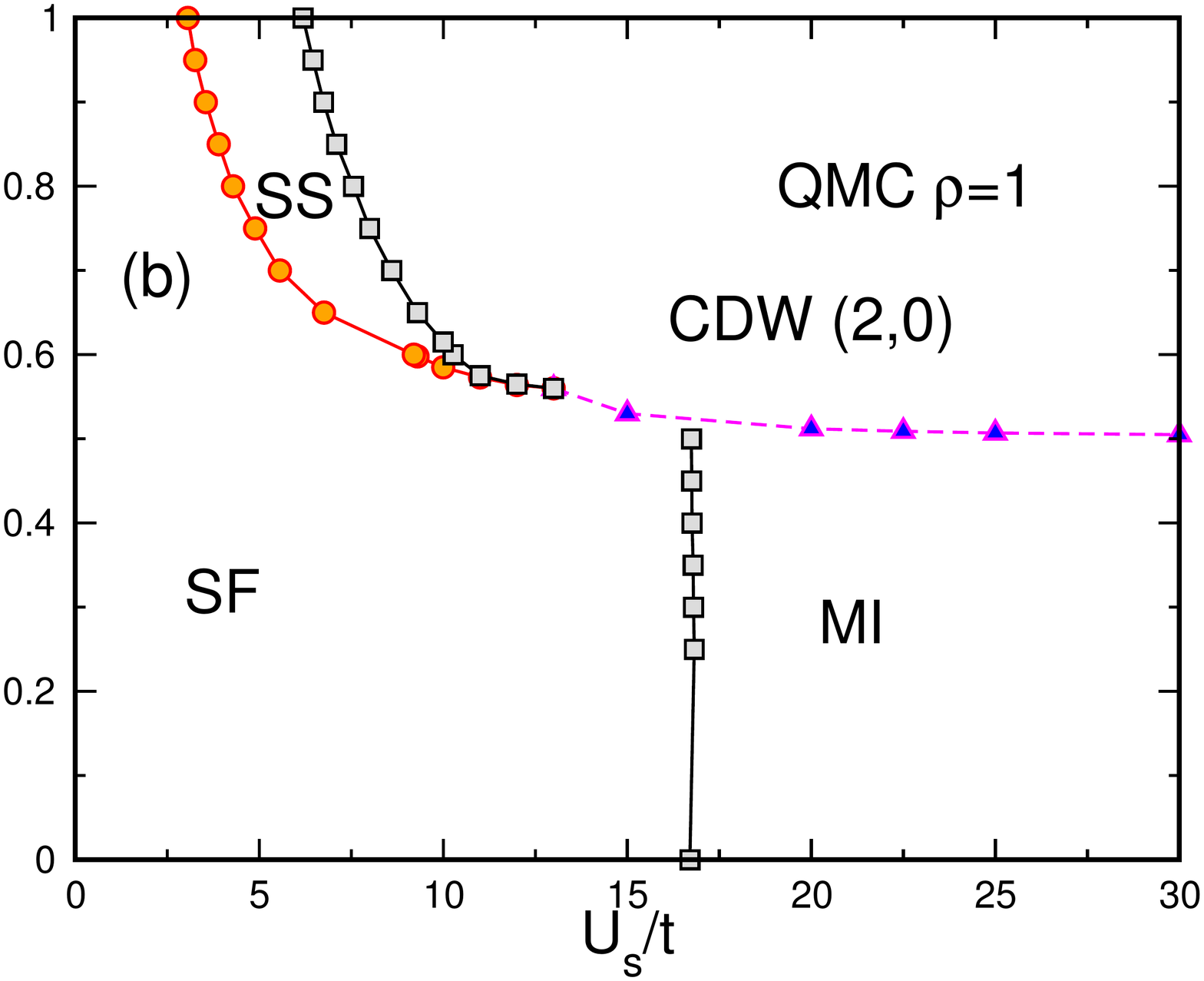}
\includegraphics[width=0.66 \columnwidth]{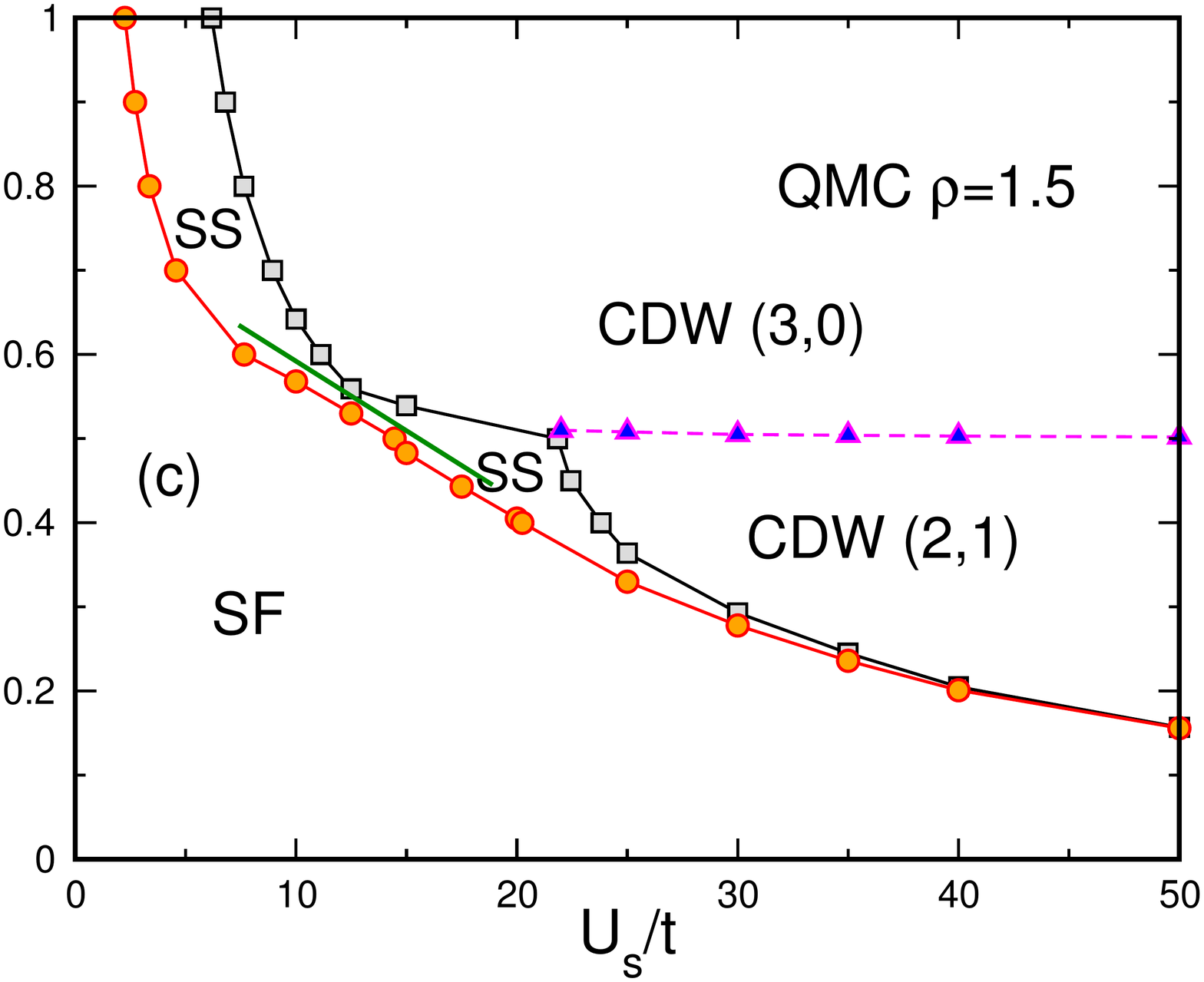}}
\caption{(Color online) Ground-state phase diagrams obtained with QMC
  simulations in the canonical ensemble for three fillings (a)
  $\rho=1/2$, (b) $\rho=1$  and  (c) $\rho=3/2$. Four phases are
  observed: superfluid (SF), Mott insulator (MI), charge density wave
  (CDW) and supersolid (SS) phases. Dashed lines indicate first-order
  transition. The black line indicates a 3D XY transition associated
  with the appearance of a condensed superfluid state. The red line
  indicates the 3D Ising transition associated with the appearance of
  density wave. At $\rho=1$, there is a small region where there is a
  direct transition between the SF and the CDW (2,0) phases where both
  the Ising $\mathbb{Z}_2$ and XY $U(1)$ symmetries are broken
  simultaneously. The green line in (c) 
 corresponds to the cut through the SS phase used in
  Fig. \ref{figcrossover} (a). 
In Sec. \ref{substab}, we will discuss
  the stability of these phases and show that, while the $\rho=1$ and
  $\rho=3/2$ SS phases are stable, at $\rho=1/2$ there is, instead, 
phase separation between a superfluid and a supersolid (SF-SS PS).
  \label{Figure_4}}
\end{figure*}
Using QMC simulations in the canonical ensemble, we determine exactly
the properties of the ground state.  We will detail our observations
in the following but we first compare the phase diagrams at fixed
fillings that we obtained with QMC (Fig. \ref{Figure_4}) and GMC
(Fig. \ref{Figure_3}). We observe the same features for all phase
diagrams, with mostly quantitative differences between the QMC and the
GMC predictions. The main difference is that the GMC technique
generally overestimates the size of supersolid or phase separation
regions.  For example,
at $\rho=1/2$ the PS region is difficult to observe below
$U_l/U_s=0.6$ as the transition lines marking the onset of density
ordering and the disappearance of condensation appear superimposed in
the limits of our simulations (Fig. \ref{Figure_4} (a)).  For
$\rho=1$, the GMC and QMC phase diagrams are essentially the same but
we observe a region where there seems to be a direct transition,
without passing via an intermediate SS phase, between the SF and CDW
(2,0) region (Fig. \ref{Figure_4} (b)).  Finally, for $\rho=3/2$, we
observe that the SS phase does not surround the two CDW phases
completely but exists for $U_l/U_s \gtrsim 0.3$ (Fig. \ref{Figure_4}
(c)).

\subsection{Phases}
\label{sec4_subA}

\begin{figure}[!h]
\includegraphics[width=\columnwidth]{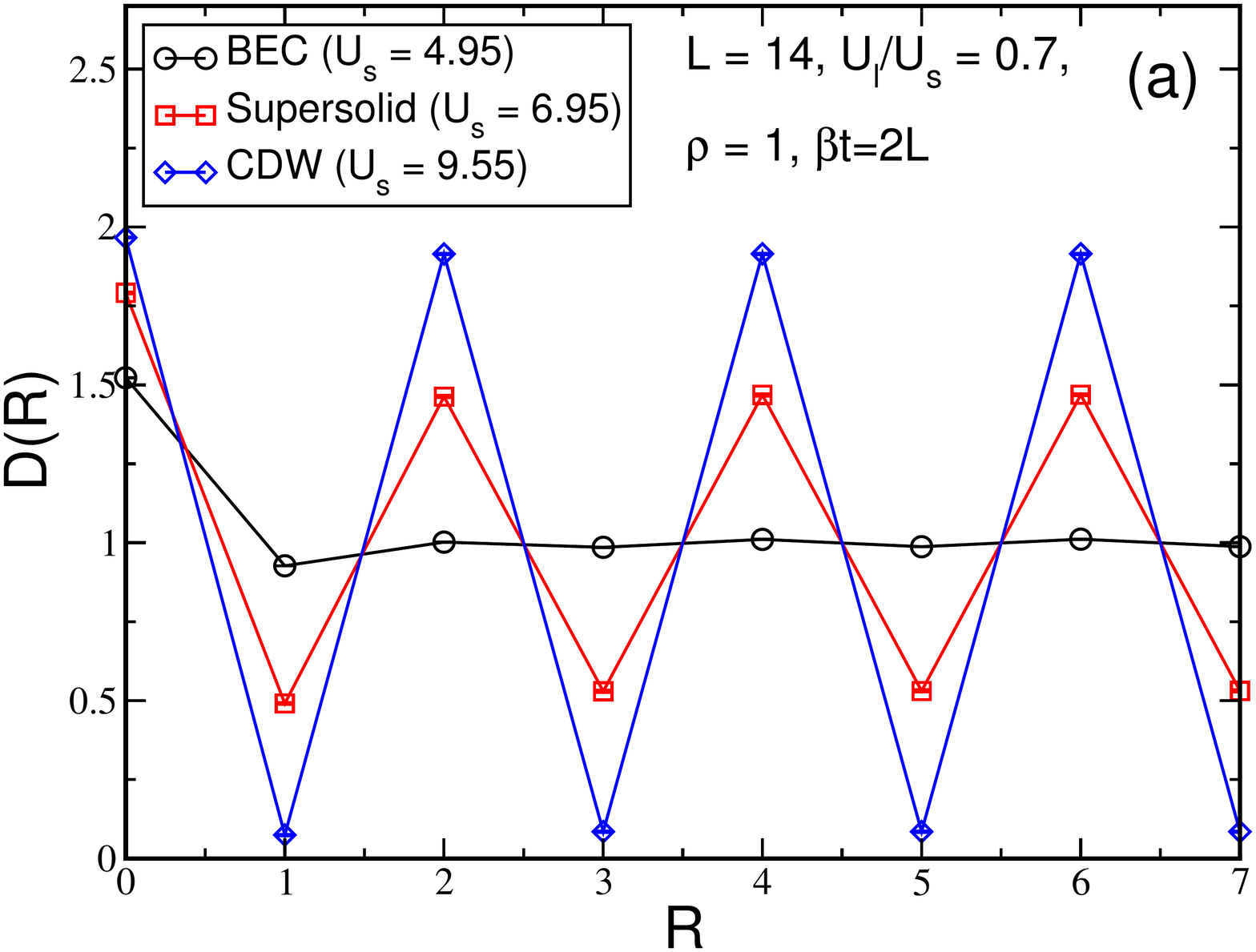}
\includegraphics[width=\columnwidth]{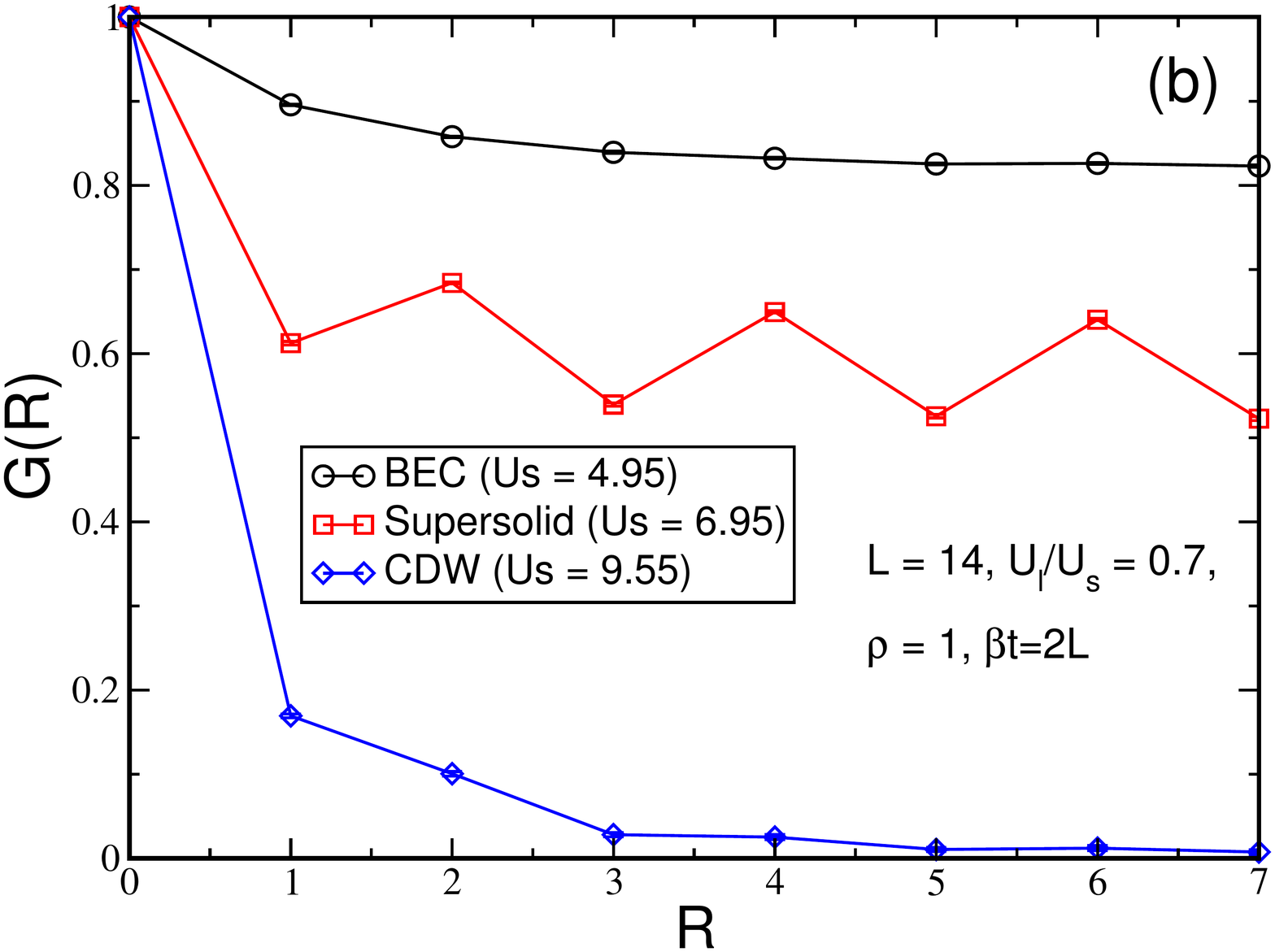}
\caption{(Color online)
The density-density correlation function $D(R)$ (a) and the Green
function $G(R)$ (b) as functions of distance $R$ in different phases
at $\rho=1$ and for a given ratio of interaction $U_l/U_s$. $D$
exhibits long range oscillations in the CDW and supersolid phases. $G$
is non zero at long distances in the supersolid and superfluid
phases. 
\label{figDG}}
\end{figure}

We first analyse the different properties of the phases using the
behaviour of the density-density correlation and Green functions
(Fig. \ref{figDG}) in the case where $\rho=1$. In the BEC or
superfluid phase, we observe long range coherence of the phase as
evidenced by the saturation at long distance in the Green function
$G({\bf R})$ value. The saturation value of $G({\bf R})$ corresponds
to the condensate fraction which is not equal to one
because of quantum depletion due to the interaction. On the contrary,
there is no diagonal order as the density correlation $D({\bf R})$
tends to a plateau with a value equal to the square of the density,
$\rho^2 =1$ in that case. The CDW phase shows strong oscillation
around $\rho^2$ at
long distance in $D({\bf R})$ while $G({\bf R})$ decays exponentially
to zero, which is characteristic of a solid phase. Finally, the
supersolid phase shows simultaneously oscillations around plateaux in
both $D({\bf R})$ and in $G({\bf R})$, which shows that both diagonal
and off-diagonal long range orders are present at the same time.  In
the Mott phase (not shown here) $G({\bf R})$ decays exponentially to
zero while $D({\bf R})$ simply tends to a plateau with a value $\rho^2$.

\begin{figure}[!h]
\includegraphics[width=\columnwidth]{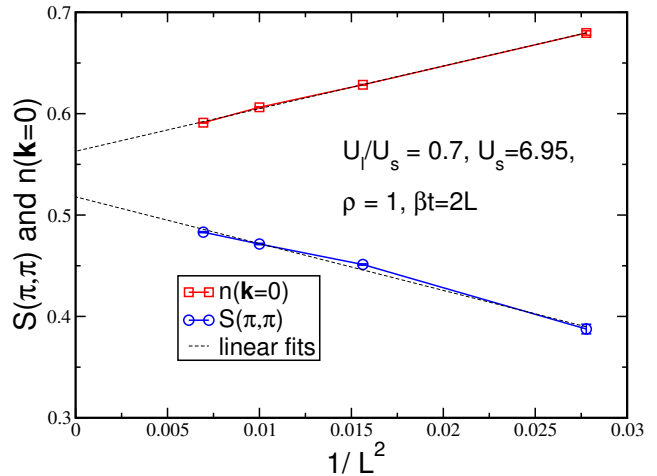}
\caption{(Color online)
The structure factor $S(\pi,\pi)$ and the condensate 
fraction $n({\bf k}=0)$ as a function of $1/L^2$ for
$\rho=1$ in the supersolid phase. Both quantities 
extrapolate to a non zero value for large sizes.
\label{figSC} }
\end{figure}

To confirm the presence of these phases, we performed finite size
scaling analyses of the structure factor, $S(\pi,\pi)$, and of the
condensate fraction, $n({\bf k}= 0)$.
  In Fig. \ref{figSC}, we show that both quantities extrapolate to non zero
values in the thermodynamic limit for values of parameters that are
chosen in the supersolid region at $\rho=1$.

\begin{figure}[h]
\includegraphics[width=\columnwidth]{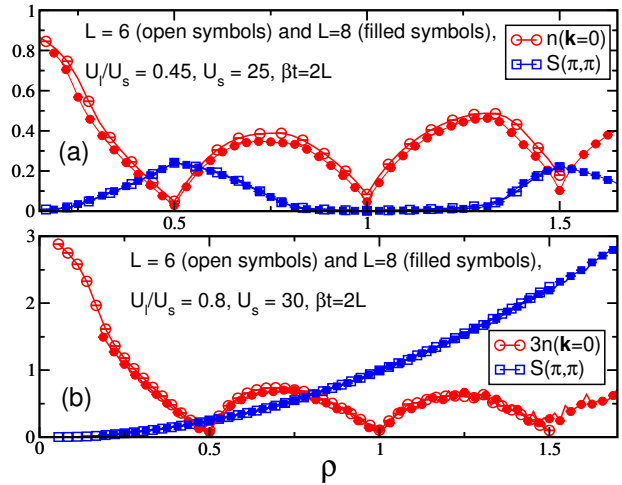}
\caption{(Color online) Cuts in the phase diagram as functions of
  $\rho$ for $U_l/U_s = 0.45$ (a) and $U_l/U_s = 0.8$ (b) and two
different system sizes $L=6$ and $L=8$. In (a) we
  see the four phases SF, CDW, MI and SS (see text). In (b) we see
  that the larger value of $U_l$ forbids the  existence of
  homogeneous phases except for low densities $\rho \lesssim 0.25$. On the
  contrary, we observe MI and SF phases in (a), i.e. regions
  where $S(\pi,\pi)$ is zero.
\label{figcut1}}
\end{figure}

As mentioned earlier, the phases observed change drastically
depending on whether $U_l$ is smaller or larger than $U_s/2$. In
Fig. \ref{figcut1} (a), we show a cut in the phase diagram, varying
$\rho$ for $U_l = 0.45\, U_s $ where we observe all four phases. For
small $\rho$, the system is superfluid; for $0.25 \lesssim \rho
\lesssim 0.75$, the system shows non zero $S(\pi,\pi)$, which means
that it is supersolid (since the condensate is still non zero) except
at $\rho=0.5$ where we find a CDW (1,0) phase. This is somewhat
surprising because supersolid phases generally do not appear in large
regions of parameter space for $\rho<1/2 \cite{Ohgoe_2012}$. This
effect is due to the infinite range interaction in the system.  For
$0.75 \lesssim \rho \lesssim 1.25$, $S(\pi,\pi)$ is zero and the
system is superfluid except at $\rho=1$ where it adopts a Mott
phase. For $1.25\lesssim\rho < 1.5$, we once again have a supersolid
and a CDW (2,1) phase at $\rho=1.5$.  On the contrary for $U_l =0.8\,
U_s$ (Fig. \ref{figcut1}(b)), we observe that, for $\rho \ge 1/2$,
there are only CDW or supersolid phases, as $S(\pi,\pi)$ is always non
zero. There is no Mott phase and the superfluid phase is limited to
the region where $\rho \lesssim 0.25$.

We remark that these results have been obtained at fixed densities
(canonical ensemble). The stability of these phases with regard to
density fluctuations, i.e. the behaviour of the system in the
grand canonical ensemble, will be discussed below (see Sec. \ref{substab}).

\subsection{Quantum Phase Transitions at fixed densities}
\label{sec4_subB}
\begin{table}[b]
\begin{tabular}{c|c|c}
  \hline
  Quantum Phase  &       symmetry      & \multirow{2}*{Type}  \\
  Transitions       &        broken     &     \\
  \hline
  \hline         
  MI-SF                 &   $U(1)$                               &   3D XY     \\
  \hline
  SS-CDW    & $U(1)$      &   3D XY       \\
  \hline
  SF-SS     &  $\mathbb{Z}_2$     &   3D Ising     \\
  \hline   
  MI-CDW   &   $\mathbb{Z}_2$      &   first-order      \\
  \hline
  SF-CDW (2,0)    &$U(1),  \mathbb{Z}_2$          &    first-order  \\
  \hline
  CDW (2,1)-CDW (3,0)  & $\O$    &   first-order    \\
  \hline
\end{tabular}
\caption{Universality classes for the quantum phase transitions of the
  phase diagrams Fig.~\ref{Figure_4}, determined using quantum Monte
  Carlo simulations.\label{tab2}}
\end{table}

\begin{figure}[!h]
\includegraphics[width=\columnwidth]{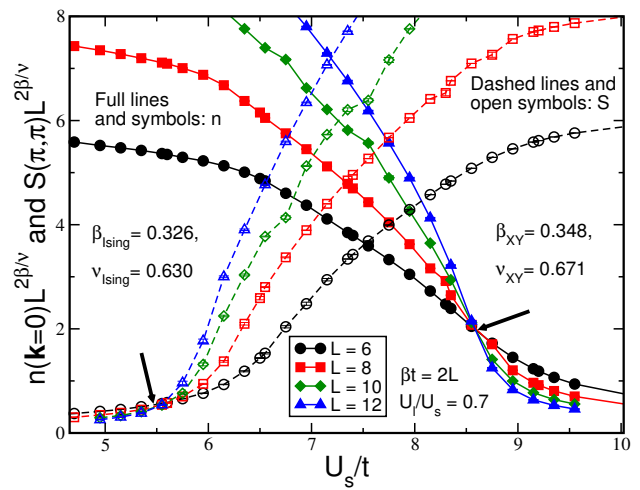}
\caption{(Color online) The transitions between the SF and SS and
  between the SS and the CDW phases for $\rho=1$.  The first
  transition is signalled by $S(\pi,\pi)$ becoming non zero (open
  symbols, dashed lines), while the second
  transition corresponds to $n({\bf k}=0)$ becoming zero (full
  symbols, full lines). In the intermediate supersolid region, both
  are non zero. The data have been rescaled with critical exponents
  $\beta_{\rm Ising}, \nu_{\rm  Ising}$ for $S(\pi,\pi)$ and $\beta_{\rm XY},
  \nu_{\rm XY}$ for $n({\bf k}=0)$, corresponding to different universality
  classes (3D XY and 3D Ising).  The arrows point to the two
  transition points where the curves obtained for different sizes
  cross each other.
  \label{figtransitionBECSSCDW}}
\end{figure}

We show here how the phase diagrams (Fig. \ref{Figure_4}) were obtained. To this
end, we analyse the quantum phase transitions between the different
phases, focusing on the $\rho=1$ case, and using finite size scaling
analysis. 
   We summarize in Table.~\ref{tab2} the different types of
phase transition that we observed. When a spatial modulation 
develops, as in the CDW or SS phases, the discrete $\mathbb{Z}_2$ translation symmetry
is broken. When the superfluidity or condensate appears, the 
continuous $U(1)$ phase symmetry is broken.
For the successive transitions
between the superfluid, supersolid and CDW phases, these $\mathbb{Z}_2$
and $U(1)$ symmetries are then broken separately or simultaneously. When
they are broken together, we expect a first order transition. When
they are broken separately, we expect 3D Ising and 3D XY universality
classes and our QMC results confirm this. We rescaled $n({\bf k}=0)$
by $L^{2\beta_{XY}/\nu_{XY}}$ and
$S(\pi,\pi)$ by $L^{2\beta_{\rm Ising}/\nu_{\rm Ising}}$. $\beta_{XY}$ and $\nu_{XY}$
here denote the critical exponents of
the 3D XY model and $\beta_{\rm Ising}$ and $\nu_{\rm Ising}$ those of the 3D Ising model. 
 With this rescaling, we expect the
curves obtained for different sizes to cross at a single point
corresponding to the transition. Increasing $U_s/t$ for a given value
of $U_l/U_s$, we observe such crossings for the SF-SS transition where
$S(\pi,\pi)$ becomes non zero and for the SS-CDW transition where
$n({\bf k}=0)$ becomes zero (see Fig. \ref{figtransitionBECSSCDW}).
We do not obtain such crossings if we use mean
field scaling exponents.

\begin{figure}[!h]
\includegraphics[width=\columnwidth]{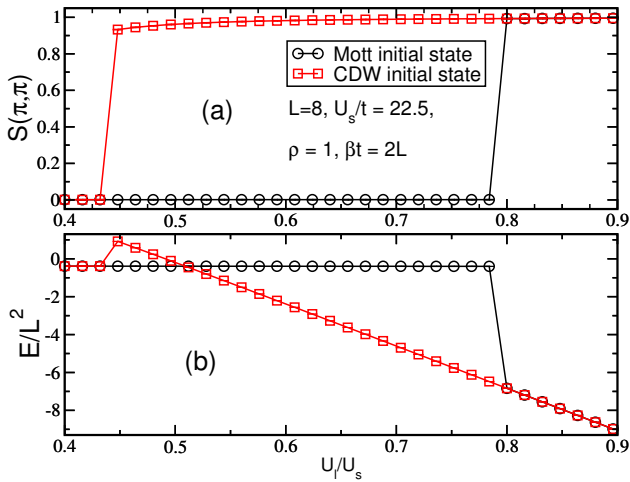}
\caption{(Color online) The structure factor (a) and ground state
  energy (b) obtained by varying $U_l/U_s$ around the MI-CDW(2,0)
  transition. The two curves in each panel correspond to different
  starting states for the QMC simulation that yield different results
  in the region where both phases coexist and the same result outside
  the coexistence region.  The transition point is located at the
  point where the two ground state energy curves meet. \label{fig8}}
\end{figure}

The transitions between different solid phases are found to be of
first order, as suggested by \cite{Esslinger_2016} for $\rho=1$.  To
confirm this, we did QMC simulations with different starting states
(CDW or MI), for a large enough system. In the region where both the
MI and CDW (2,0) phases coexist, the system remains in the local
energy minima corresponding to the phase with which we started the
simulation.  An hysteresis effect is then observed for $S(\pi,\pi)$ as
$U_l/U_s$ is varied (Fig. \ref{fig8}(a)).  In the coexistence region
around the transition we obtain two different values of $S(\pi,\pi)$
while the two initial conditions yield the same values when we are far
away from the transition point and no longer have coexistence of
metastable and stable states.

Examining the ground state energy, i.e. the free energy,
measured with both initial states, the transition point is located where
the two energy curves cross. Below this crossing point, the CDW state
is metastable and the Mott state is stable (Fig. \ref{fig8}(b)), while
the opposite is true above this point.

We observe the same hysteresis effects at $\rho=1$ in the small region
($13<U_s/t<17$ in Fig. \ref{Figure_4}) between the SF and CDW
phases. This confirms the presence of a direct first order transition
from the SF to the CDW (2,0) in this region and the absence of a
supersolid which then only exists for $U_s/t<13$.

As the interaction $U_s/t$ grows, for $\rho=1/2$ and $\rho=3/2$,
the PS and SS regions becomes narrower and eventually disappears. There is then 
a direct transition between SF and CDW (1,0) at $\rho=1/2$ and between
SF and CDW (2,1) at $\rho=3/2$. We were not able to obtain 
a definite conclusion concerning the nature of these transitions,
due to the difficulty of the QMC simulations in this large interaction regime.

\begin{figure}[!h]
\includegraphics[width=\columnwidth]{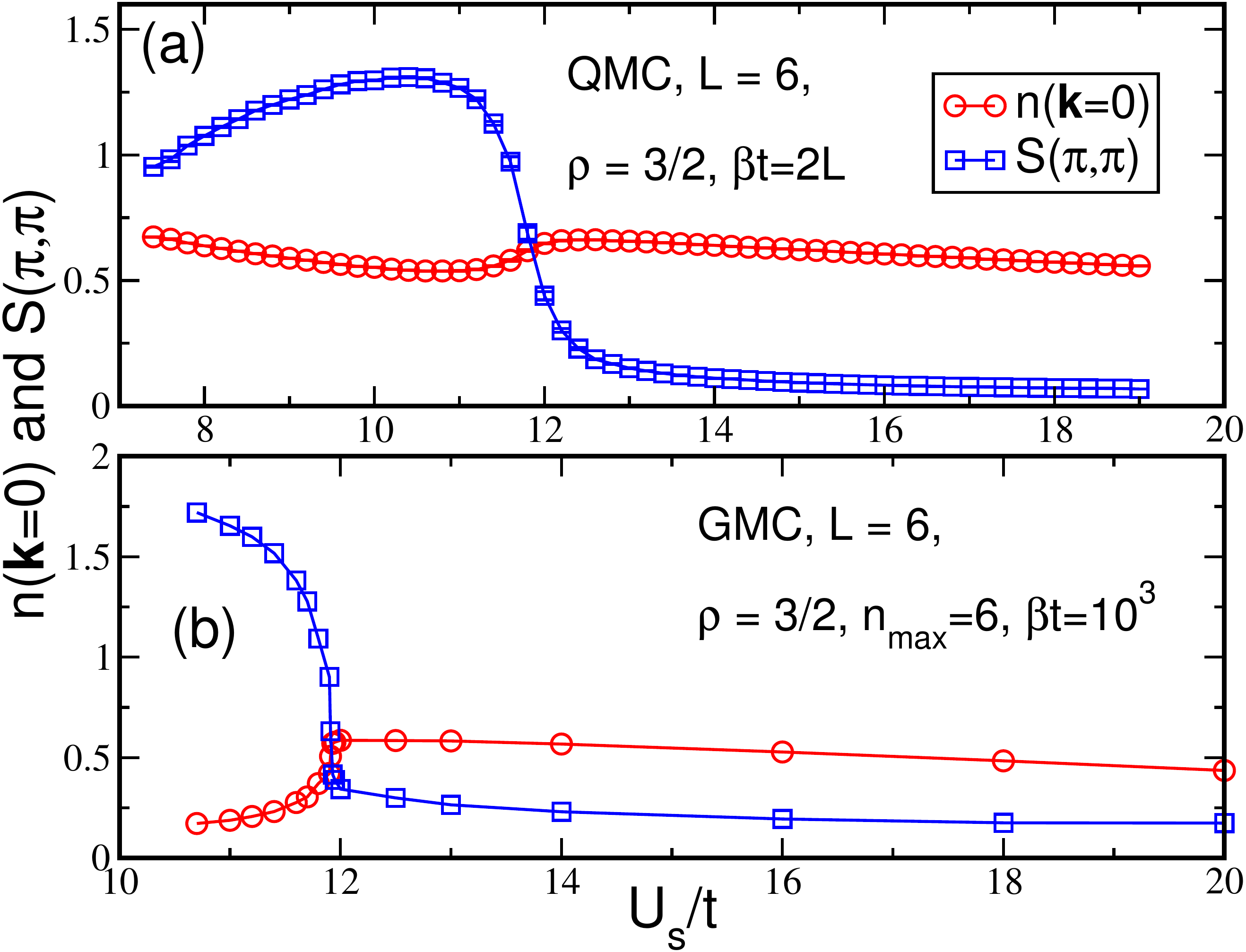}
\caption{(Color online) Cuts along lines inside the $\rho=3/2$
  supersolid phase (see the green lines in Figs. \ref{Figure_3}(c) and
  \ref{Figure_4}(c)) 
   made with QMC (a) and GMC (b). In both cases, we
  observe a crossover between two different regimes. As $U_s/t$ grows
  ($U_l/U_s$ diminishes) the SS goes from a regime where
  $S(\pi,\pi)$ is large, due to the proximity with the CDW (3,0)
  phase, to a regime where it is much smaller, when the SS is close to
  the CDW (2,1). As the crossover between these two SS regimes happens
  (around $U_s/t = 12$), the condensate fraction $n({\bf k}=0)$ increases.
  \label{figcrossover}}
\end{figure}

For $\rho=3/2$, we also observe a first order transition between the
CDW (2,1) and the CDW (3,0) phases. This is somewhat surprising since
the same spatial symmetry is broken in both phases and one might have
expected a crossover. The proximity of this first order transition to
the narrowing of the $\rho=3/2$ SS phase (see Fig. \ref{Figure_4}(c))
raises the question of a possible transition between two different SS
phases. Cutting along a line through the SS phase (see
Fig. \ref{Figure_4}(c)), we do not find a transition but a crossover between two
different regimes (Fig. \ref{figcrossover}(a)). Close to the CDW (3,0)
phase ($U_s/t < 12$), the oscillations of density are much larger in
the SS phase, which is marked by larger value of $S(\pi,\pi)$.
$S(\pi,\pi)$ becomes much smaller close to the CDW (2,1) phase
($U_s/t>12$). Such a large change is expected as, in the large
interaction limit, $S(\pi,\pi)$ goes from 9/4 down to 1/4 between the
CDW (3,0) and (2,1) phases.  While $n({\bf k}=0)$ generally decreases
with increasing $U_s/t$, it increases slightly in the crossover region
($U_s/t \simeq 12$). We also observe this behaviour with the GMC
approach (Fig. \ref{figcrossover}(b)).

\subsection{Stability of phases\label{substab}}

So far, we mainly discussed the phases and transitions that appear at
a constant particle number and mentioned that, in some cases, 
Hubbard systems can be unstable to
phase separation as the density is allowed to
fluctuate. In particular, it was shown that such a phase separation
can occur between superfluid and solid phases and be mistaken for a
stable supersolid phase \cite{batrouni2000}.  This is especially
important in the present case as long range attractive interactions,
if strong enough, tend to destabilize the system and lead to a
collapse of the particles (for example for $U_l/U_s>1$ in the MF description) or to the
presence of metastable states \cite{menotti2007}.

We, therefore, analyse the stability of phases by studying the
evolution of the density, $\rho$, as a function of the chemical
potential $\mu$.  In the grand canonical ensemble $\rho$ is calculated
directly as an average for a given value of $\mu$. In the canonical
ensemble, $\rho$ is fixed and $\mu$ can be calculated, for $kT\simeq
0$, as $\mu(N)=E(N+1)-E(N)$ where $E(N)$ is the ground state energy of
the system with $N$ particles.  In the GCE, an unstable region is
characterized by a jump in the density as $\mu$ is varied. The
intermediate densities do not correspond to a stable phase. Using CE,
one can choose a filling corresponding to these intermediate densities
but then the region is signalled by a curve $\rho(\mu)$ with a
negative slope\cite{batrouni2000}.

\begin{figure}[h]
\includegraphics[width=\columnwidth]{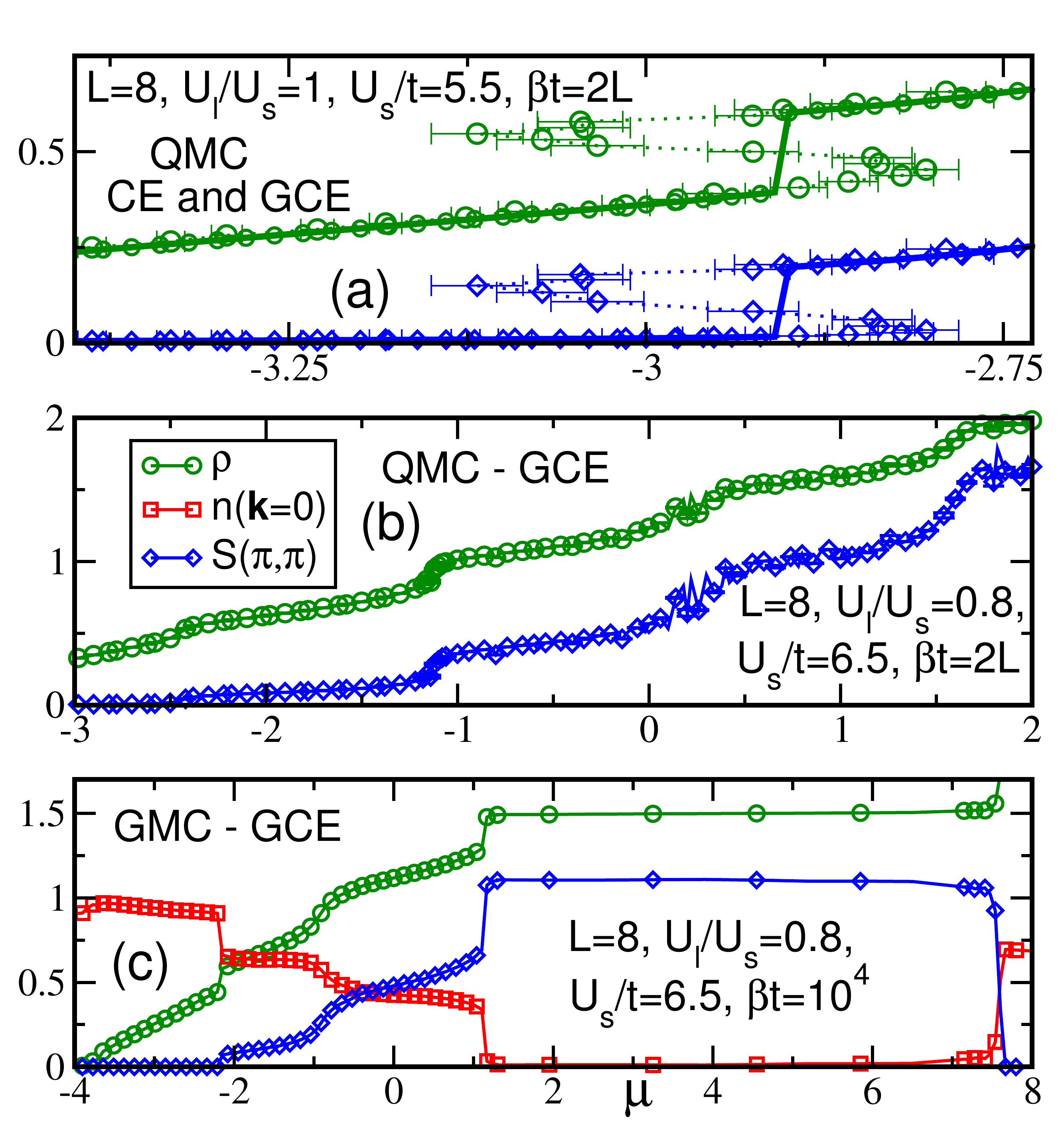}
\caption{(Color online) $\rho$, $n({\bf k}=0)$ and $S(\pi,\pi)$ vs
  $\mu$. Dotted lines correspond to canonical simulations while
  continuous lines correspond to grand canonical ones.  (a) Using QMC
  CE simulations, we observe a region in $\rho(\mu)$ with negative
  slope around $\rho=1/2$ indicating instability. $n({\bf k}=0)$ is
  always non zero (not shown to keep the figure uncluttered).  There
  is, therefore, a discontinuous transition from a superfluid for $\rho
  \lesssim 0.39$ to a supersolid for $\rho \gtrsim 0.60$ which is also
  signalled by a jump in $\rho(\mu)$ when using the GCE.  There is no
  stable phase at $\rho=0.5$ for these parameters.  (b) For a
  different set of parameters, we do not observe jumps in the
  $\rho(\mu)$ curve (QMC GCE simulations) which shows that the
  supersolid phases at $\rho=1$ and $\rho=1.5$ are stable.  $n({\bf
    k}=0)$ is always non zero (not shown).  (c) Similar behavior is
  observed using the Gutzwiller approximation.   In
  the case presented here we observe a direct transition from SF at
  $\rho \lesssim 0.44$ to SS for $\rho \gtrsim 0.58$ and a
  discontinuous transition from SS for $\rho \lesssim 1.29$ to
  solid phase for $\rho=1.5$. We then have two density regions where
  the system is unstable. In (b) and (c) $S(\pi,\pi)$ has been
  multiplied by 0.5 to improve visibility.
  \label{instab}}
\end{figure}

We found that some regions of the phase diagrams are unstable. For
example doping around $\rho=0.5$ for $U_s/t = 5.5$ and $U_l/U_s=1$
(Fig. \ref{instab}(a)), we observe a large unstable region which
encompasses $\rho=0.5$ and leads directly from a superfluid phase for
$\rho \lesssim 0.39$ to a supersolid phase for $\rho \gtrsim
0.60$. This is attested by the behavior of $\rho(\mu)$: a jump is
observed in the GCE simulations and a corresponding negative slope
region is observed in CE simulations.  This means that what appears to
be a supersolid phase with the canonical simulations for $\rho=0.5$ in
the phase diagram Fig. \ref{Figure_4}(a) is, 
as mentioned earlier, not a stable phase but
corresponds to a region of separation between SF and SS phases
(SF-SS PS region in Fig. \ref{Figure_4}(a)).

On the other hand, cutting through the phase diagram at
$U_s/t=6.5$ and $U_l/U_s=0.8$ (Fig. \ref{instab}(b)), i.e. going
through the SS regions at $\rho=1$ and $\rho=1.5$ (see
Fig. \ref{Figure_4}(b) and (c)), we do not observe jumps in
$\rho(\mu)$ using GCE simulation (Fig. \ref{instab}(b)) and conclude
that the supersolid phases are in fact stable for these densities, with these parameters.
Finally, we performed some simulations (not shown here) for large
values of the interactions ($U_s/t = 30, U_l/U_s = 0.8$) where we
observed that the only stable phases are the solid phases obtained for
integer or half-integer densities: there is no stable superfluid or
supersolid phases for intermediate densities.

This question of stability can also be addressed within the Gutzwiller
approximation.  In Fig. \ref{instab}(c), we observe jumps in
$\rho(\mu)$ that mark the presence of unstable regions. In particular,
again observe that the system is not stable for $\rho=0.5$ and that
there is a SF-SS discontinuous transition around that density.  The
parameters used for Fig. \ref{instab}(b) and (c) are the same but, as
the phase diagrams Fig. \ref{Figure_3} and Fig. \ref{Figure_4} are
slightly different, the phases that are present in these two cuts are
different. For example, at $\rho=1.5$, we find a SS phase in the QMC
simulations while we have a solid phase in the GMC simulations.

\section{Conclusions}
\label{sec_5}

\begin{figure}
\includegraphics[width=\columnwidth]{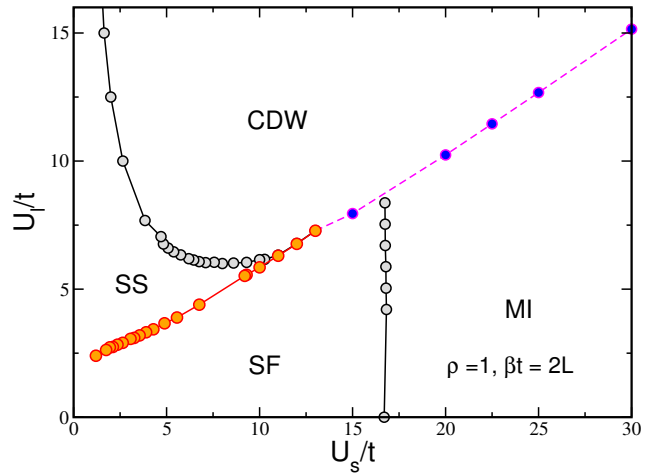}
\caption{(Color online) The phase diagram (Fig.~\ref{Figure_4}~(b))
  for $\rho=1$ rescaled for a direct comparison with experimental data
  (figure 2 of extended data \cite{Esslinger_2016}). Compared to the
  experiment, the  superfluid and supersolid regions are found to be
  smaller. 
 \label{RPD}}
\end{figure}

Using exact quantum Monte Carlo and approximate mean field
calculations, we studied the phase diagram of a bosonic Hubbard model
with infinite range interactions which has been proposed to describe
recent experimental results of the ETH-Zurich
group\cite{Esslinger_2016}.  Our results confirm that the model
correctly captures the essential physics of the experiments. In
particular, we confirm the existence of a supersolid phase and also
the nature of the phase transitions, especially the first order
transition between the MI and CDW at $\rho=1$. We observe a small
region where there is, at $\rho=1$, a direct first order SF to CDW
transition that was not experimentally observed.  

There remains, however, quantitative differences in the extent of the
different phases. This is easier to discuss by comparing the phase
diagram Fig. \ref{RPD}, represented in the plane $(U_s/t,U_l/t)$, with
the one provided as Figure 2 of extended data in
Ref.~\onlinecite{Esslinger_2016}. In the experimental data, the SF
phase is observed up to $U_s/t \simeq 25$ whereas we observe a
transition around $U_s/t \simeq 17$.  The SS region is also much
larger in the experimental figure extending up to $U_s/t \simeq 30$
and $U_l/t \simeq 25$ whereas we observe it in the region where
$U_s/t$ and $U_l/t$ are smaller than 10.  We believe these
discrepancies to be due to the fact that we performed bulk
simulations (i.e. with no trap) whereas the experiment takes
place in a harmonic trap. However, the main point is that the
existence of the supersolid phase observed in the experiment is
confirmed. It is a true thermodynamically stable phase and not a
simple mixture of superfluid and solid phases.

We also extended the phase diagram to other densities and confirmed
predictions that were made using mean-field calculations
\cite{Donner_2016, Mueller_2016}.  We also discussed the stability of
the observed phases with respect to phase separation and found that
the $\rho=1$ and $\rho=1.5$ supersolid phases are stable whereas
there is phase separation between a SF and a SS for $\rho=1/2$. 

\begin{acknowledgments}
We thank T. Roscilde for constructive discussion on the GMC technique.
We thank G. Morigi and T. Donner for useful discussions.  This work
is supported by the Alexander von Humboldt-Foundation.  Some
calculations have been performed on the PSMN centre of the ENS-Lyon.
\end{acknowledgments}

\end{document}